\documentclass[a4paper,english]{revtex4}
\usepackage[T1]{fontenc}
\usepackage[utf8]{inputenc}
\usepackage{color}
\usepackage{amstext}
\usepackage{graphicx}

\makeatletter

\providecommand{\tabularnewline}{\\}

\makeatother

\usepackage{babel}

\begin{document}
\title{A Preonic Model with Colour - Spin Symmetry}

\author{Oktay Doğangün}
\email{oktay.dogangun@cern.ch}
\affiliation{Department of Physical Sciences, University of Naples \& INFN,
Naples, Italy.}

\author{Saleh Sultansoy}
\email{ssultansoy@etu.edu.tr}
\affiliation{Physics Department, TOBB ETU, Ankara, Turkey}
\affiliation{Academy of Sciences, Institute of Physics, Baku, Azerbaijan}

\author{N. Gökhan Ünel}
\email{gokhan.unel@cern.ch}
\affiliation{University of California at Irvine, Physics Department,Irvine, CA,
USA}

\begin{abstract}
We have constructed a preonic model starting from a coloured fermionic
preon and by postulating a new symmetry, MUSY. This new symmetry is
defined via the MU number involving colour, charge and spin properties
of the preons. We show that all the known fields of the Standard Model
(SM) can be constructed using the fermionic preon and 6 preonic scalars,
its MUSY partners. As an example, we present well known $\beta$-decay
process at MUSY level. MUSY also forbids some processes such as proton
decay (preserving the baryon number) and hence, it is compatible with
current experimental results. In this model the number of SM generations
arises to be three naturally.

The MUSY generalization of the SUSY algebra is constructed and the
MUSY invariant Lagrangian is also explicitly written. Similar to other
preonic and supersymmetric models, a number of new particles are predicted.
These particles do not interact with any of the SM fermions but only
with the gauge bosons. These particles could be dark matter candidates. 
\end{abstract}

\maketitle

\section{Introduction}

Although the Standard Model (SM) of particle physics is able to explain
the experimental results obtained so far \cite{PDG-1}, its problems
and open questions shed some doubt on its prospects as being the ultimate
theory of fundamental particles and their interactions. The large
number of free parameters in the model, the nature of the elementary
particles, the reason for their observed mass hierarchy, the electroweak
symmetry breaking mechanism are some examples of these open issues.
A number of alternative theories have been proposed as the cure to
these problems. For example, Grand Unified Models \cite{GUT-1}, Technicolour
\cite{Technicolour-1}, Compositeness \cite{Compositeness-1} and Supersymmetry
\cite{SUSY-1} can be cited. The composite models are particularly
interesting for the continued simplification they offer. Similar situations
have been seen at least twice in the the past: The big and complex
periodic table of elements has been understood in terms of three constituents
and their combinations: the proton, the neutron and the electron.
Similarly, to solve the hadron inflation of the 1960ies, the quark
model had to be introduced. The hope of the authors is to propose 
a solution to some of the aforementioned problems with an effective model
at the preonic level. Considering elementary particle
family replication and especially the fermion mixings
as hints for preonic structure of some of SM particles, a number of
preonic models, with different levels of compositeness, have been
studied by various groups. Below we list a few examples.

H. Harari \cite{harari} and A. M. Shupe \cite{shupe} proposed two
coloured fermionic preons (T,V) with which the first generation of
SM quarks and leptons can be formulated as three preon bound states.
In this model, the other fermion generations are thought to be excitations
of the first generation and the SM bosons are assumed to be fundamental
particles. Considering Harari-Shupe model, Buchmann and Schmid \cite{BuSch-1}
later proposed some quark-like structures with hypercolours so that
the model become compatible with 't Hooft anomaly. 

H. Fritzsch and G. Mandelbaum considered two fermionic and two bosonic
preons ($\alpha,\beta,x,y$) to construct all SM fundamental particles
except the photon and gluons \cite{FriztschM-1}. In this model, the
first generation of SM quarks and leptons are composed of a fermionic
and a bosonic preon, while $W$, $Z$ and Higgs bosons are assumed
to consist of a fermionic preon and its anti-particle. The other generations
are constructed by adding one or more {}``gluons'' of the preon
binding force (haplodynamics) to the appropriate state. Therefore,
this model is not able to give any constraint on the number of the
generations in the SM.

J. Dugne, S. Fredriksson, J. Hansson and E. Predazzi, in their Preon
Trinity Model \cite{Trinity-1}, proposed three coloured fermionic
preons ($\alpha,\beta,\delta$) with which they construct three scalar
bound states ($x,y,z$). With the bound state of a fermionic and a
scalar preon (or its anti-particle) they construct all SM fermions
and more: one additional charged lepton, two neutral leptons and three
additional quarks emerge naturally from the model. With the various
combinations of the fermionic preons, they construct the electroweak
gauge bosons and more: the model predicts two additional charged bosons
and four additional neutral ones. 

A. Celikel, M. Kantar and S. Sultansoy \cite{Celikel} proposed two
scalar and two fermionic preons, all coloured. In this model leptons
are bound states of one fermionic preon and one scalar antipreon whereas
antiquarks consist of one fermionic and one scalar preon. As a result
each SM lepton has its colour octet partner and each SM antiquark
has its colour sextet partner.

S. Ishida and Sekiguchi \cite{ishida} have discussed a left-right symmetric
preon model assuming the quarks, leptons and weak bosons as composite states 
and the weak interaction is a secondary effective one without the need of
breaking a
fundamental symmetry. The fermions are formed by a fermionic and a bosonic
preons where 
the fermionic preons are colour singlet but isospin doublet and bosonic preons
are
$\bar{\bf 3} \oplus {\bf 1}$ colour. The conservation of baryon and lepton
number is
included by hand.

The goal of this work is to use the key concepts from compositeness
and supersymmetry ideas for constructing a new model with charge-colour
and spin symmetry at preonic level. Using this preonic model, we attempt
to reduce the number of free parameters, answer the questions on the
number of elementary particle families, to propose a candidate for the Dark
Matter (DM) and to solve the hierarchy problem. It will be shown that
all three fermion generations of the SM can be obtained from a single
chiral coloured preon using appropriate symmetries. Using current-current
interactions, the known bosons of the SM naturally emerge in the right form,
e.g. gluon masses vanishingly small. The baryon and lepton number
conservations are emergent from the model itself,
prohibiting the anomalies like the proton decay.

The rest of this manuscript is organized as follows: in the next section,
the reader will be motivated for a colour and spin symmetric preonic
model, MUSY model, which defines all known SM fundamental particles
as preonic bound states. The section \ref{sec:Dynamics-of-MUSY}
will go into the dynamics of the MUSY model to define the MUSY algebra
and to construct the effective MUSY Lagrangian. In Section
\ref{sec:SMparticles}, 
the preonic bound states forming SM particles aer discussed.
Section \ref{sec:fenomenoloji} contains a MUSY description
of the well known processes of proton and $\beta$-decay followed by an
assesment of the CKM and MNS mixings in MUSY model.
In section \ref{sec:Predictions-of-MUSY},
a number of testable MUSY predictions are considered: excited fermions and
dark matter candidates. The last section contains some discussions
on the phenomenology of the MUSY model.

\section{Building the MUSY model\label{sec:Building-the-MUSY}}
The construction of the model starts with a single fermionic chiral
preon colour triplet of charge $-e/6$. Firstly, we define a conserved
{}``MU''%
\footnote{The word {}``mu'' means {}``mystery, yet unknown, hidden'' in
Old Turkic (and in some Altaic languages). Today the word is living
in Modern Turkish as interrogative.%
} number for each preon as follows: \begin{equation}
MU=QC+S+H\,,\label{eq:MUnumber}\end{equation}
where Q is electrical charge, C is number of colours (3 for colour,
-3 for anticolour triplet, 1 for colour singlet), S is spin and H
is helicity ( $+1/2$ for right, $-1/2$ for left components). The
conserved quantity MU will arise from a transformation which makes
the action invariant, i.e., a MU symmetry or MUSY in short.

\begin{table}[ht]
\caption{Preons of the MUSY model: musions. The subscript $i$ runs from 1
to 3 and $a$ runs for three colour charges (Red, Green and Blue). 
\label{Tab:MUSY-preons} }
\begin{tabular}{l|cccc|c}
 & Q  & C  & S  & H  & MU \tabularnewline
\hline
\hline 
$\psi_{a}$  & $-1/6$  & \textbf{3}  & $1/2$  & $+1/2$  & $+1/2$ \tabularnewline
$\bar{\psi}^{a}$ & $-1/6$  & \textbf{3}  & $1/2$  & $-1/2$  &
$-1/2$\tabularnewline
$\phi_{i}$ & $+1/2$  & \textbf{1}  & $0$  & $0$  & $+1/2$\tabularnewline
$\hat{\phi}_{i}$ & $-1/2$  & \textbf{1}  & $0$  & $0$  & $-1/2$\tabularnewline
\hline
\end{tabular}
\end{table}

A symmetry transformation that changes spin-1/2 to spin-0, colour
triplet to singlet (or vice versa) leads us to generate six scalar
preons starting from the single initial fermionic preon with left
and right states. Therefore, each helicity state of the fermionic
preon will give three complex scalars since it is a colour triplet
as well as a complex spinor field. According to these constraints,
one may write the MU transformation mappings as follows using the 
notation of \cite{Martin}:

\begin{eqnarray}
\label{eq:MUSY-transformation}
\begin{array}{rcl}
\psi_{\alpha, a} & \rightarrow & \phi_{i}\label{eq:aksi2-1}\\
\bar{\psi}^{\dot{\alpha}, a} & \rightarrow & \hat{\phi}_{i}^{*}\\
\phi_{i} & \rightarrow & \psi_{\alpha, a}\\
\hat{\phi}_{i} & \rightarrow & \bar{\psi}^{\dot\alpha, a}
\end{array}
\end{eqnarray}
where $\psi_{\alpha, a}$ and $\bar{\psi}^{\dot\alpha, a}$ are respectively left
and
Hermitian conjugate of the right handed fermionic preons with colour index
$a=R,G,B$, and the
fields $\phi_{i}$ and $\hat{\phi}_{i}$ denote the scalar preons
with family index $i=1,2,3$. The quantum numbers of the MUSY preons are
presented
in Table \ref{Tab:MUSY-preons}. One may notice that the transformation
maps colours to families in addition to mapping fermions to bosons
as in supersymmetry \cite{Martin}. This property makes the model look like a
``colour-family
extension of supersymmetry'', that is, MUSY will reduce to a SUSY
if one equates the colour indices to the fermion family indices,
which means setting $a=i$ \cite{WessZumino}.

In supersymmetry, a fermionic parameter, denoted by $\epsilon$,
was introduced to suppress the spinor indices in one side of the supersymmetric
transformations
\cite{SUSY-1}. In MU symmetry, as one may notice, there has
to be a similar variable that would suppress both colour and family indices.
Hence, a MU symmetric parameter, denoted by $\zeta_{i}^{a}$, is to
be introduced. It will lead to CKM mixing of quarks and electroweak
symmetry breaking as it will be discussed later in Section
\ref{sec:fenomenoloji}. Notice that the reduction
of MUSY to SUSY is then nothing but setting the parameter $\zeta_{ia}$
proportional to Kronecker delta $\delta_{ia}$. The complete transformation
will be discussed in Section \ref{sub:MUSY-transformation}.

\section{Dynamics of the MUSY model\label{sec:Dynamics-of-MUSY}}

Generally, most of the preonic models suffer from the lack of a convincing
dynamical
explanation for the problem of the bound states consisting of both fermion
and scalar preons. The crucial problem in those models arises from
the absence of an explicit interaction between preons. Some models constist of
a Yukawa interaction in which the preons have a common gauge symmetry.

\begin{figure}[ht]
\includegraphics[scale=0.5]{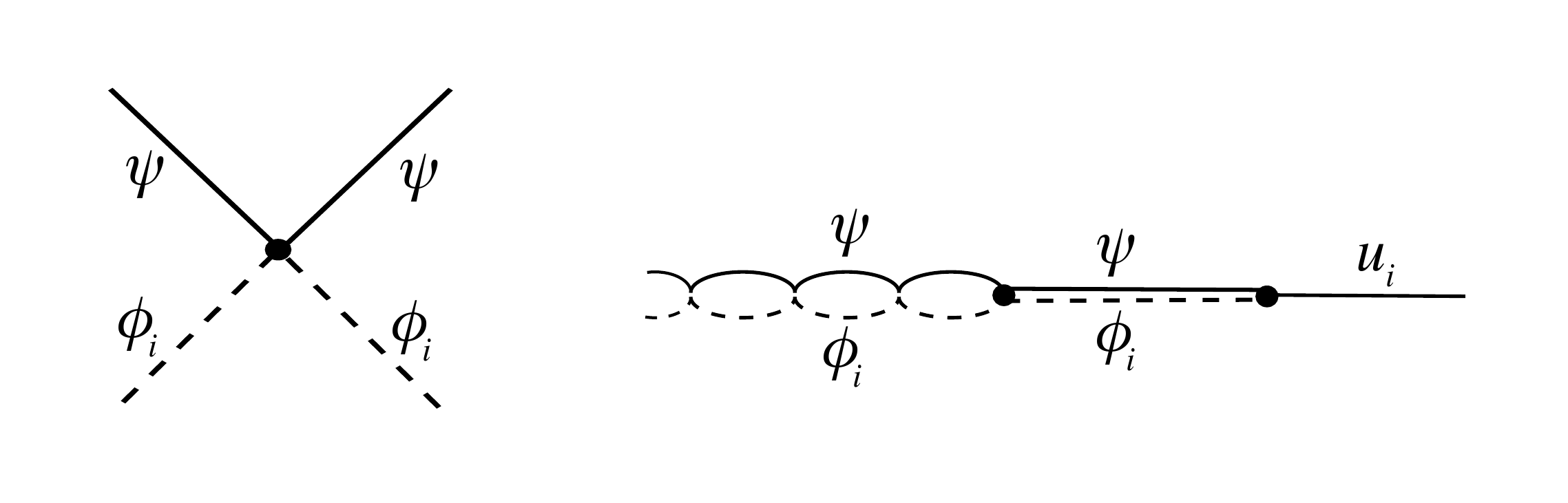}
\caption{Vertex of fermionic and scalar preons (left). 
Bound state of a fermionic preon and a scalar preon forming a quark (right).
\label{fig:boundstates}}
\end{figure}

In MUSY model, there does not exist a Yukawa interaction between scalar preons
and fermionic preons because of the chirality, i.e., a Yukawa term is not
invariant
under $SU(3)_L \otimes SU(3)_R$ or $SU(2)_L \otimes SU(2)_R$ global symmetry for
which the scalar sector and the fermionic
sector remains invariant, respectively. However, the MUSY 
model has an advantage of having Fermi-like interactions which
lead one to have an approximate gauge symmetry as will be discussed in the next
sections.
These approximate gauge interaction terms, here denoted by ${\cal L}_{int}$,
will effectively generate supersymmetric Noether currents \cite{Iorio:2000xv} of
preons in
low energies (i.e., SM scale) as follows:
\begin{eqnarray}
 {\cal L}_{\mbox{eff}} &=& - \frac{i}{2} \int d {\bf x} \, T \, \lbrace {\cal
L}_{int} ({\bf x}), {\cal L}_{int} ({\bf x}^\prime) \rbrace \\
&=& \frac{G}{\sqrt2} \, \bar\zeta J_N^\mu \bar\zeta J_{N\mu} + \frac{G}{\sqrt2}
{\bar J}_N^\mu \zeta {\bar J}_{N\mu} \zeta + \mbox{Higher dimensional terms}
\end{eqnarray}
where it is used the Wilson short distance method \cite{wilson}, 
\begin{eqnarray}
\label{eq:noether_current}
J_{N}^{\mu} & = & \gamma^{\nu}\gamma^{\mu}\Psi\partial_{\nu}\Phi^{\dagger}
\end{eqnarray}
is a fermionic current and have both colour and family indices, $\zeta$ is the
MUSY parameter, $\gamma^{\mu}$ are Dirac matrices, $\Psi$ is the four-component
fermionic preon, $\Phi$ is the scalar preon doublet.
The $\left( \Phi \partial \Psi \right)^2$ vertex occured in the effective
Lagrangian of the MUSY model is shown on the left of Figure
\ref{fig:boundstates}.
On the right of the figure, it is described the low energy transition of the
bound state schematically.

In this Section it will be described the dynamics of the MUSY model in the
following picture:
\begin{eqnarray}
 {\cal L}_{\mbox{preon}} \left( \Psi, \Phi \right) \rightarrow {\cal
L}_{\mbox{app.}} \left( \Psi, \Phi, {\bf G} , {\bf W} , {B} \right)
\rightarrow {\cal L}_{\mbox{eff}} \left( q, \ell, {\bf G} , {\bf W} , {
B}\right)
+ {\cal L}_{\mbox{DM}}
\end{eqnarray}
where $\Psi, \Phi$ are preons, $q, \ell$ are quarks and leptons, ${\bf G} , {\bf
W} , {B}$ are vector bosons with approximate guage symmetry and $ {\cal
L}_{\mbox{DM}}$ is the Lagrangian of the composite fields other than SM fields.

\subsection{MUSY transformation\label{sub:MUSY-transformation}}

MUSY transformation maps a colour-triplet Weyl spinor to three scalar
fields, and vice versa, since bosonic number of degrees of freedom
($n_{B}$) and fermionic number of degrees of freedom ($n_{F}$) should
be equal in order to handle the naturalness problem
of scalar preons, that is, when $n_{F}=n_{B}$, the huge mass correction
to a scalar field vanishes. Indeed, a (complex) colour-triplet Weyl
spinor has 12 degrees of freedom and each (complex) scalar have 2
degrees of freedom. The MU transformation also conserves
the number of degrees of freedom between bosonic preons and fermionic
preons as shown in Table \ref{tab:ndf_musy}. 

\begin{table}[ht]
\caption{Fermionic and bosonic numbers of degrees of freedom (ndf) in MUSY
model. The order of the product in ndf row is colour $\times$ family $\times$
spinor. \label{tab:ndf_musy}}
\begin{tabular}{c|c||c||c|c||c||c|c|c|c|c}
\multicolumn{1}{c|}{} & \multicolumn{6}{c|}{fermions} &
\multicolumn{4}{c}{bosons}\tabularnewline
\hline
\hline 
\multicolumn{1}{c|}{Fields} & \multicolumn{3}{c|}{$\psi$} &
\multicolumn{3}{c|}{$\bar{\psi}$} & $\phi_{i}$ & $\phi_{i}^{*}$ &
$\hat{\phi}_{i}$ & $\hat{\phi}_{i}^{*}$\tabularnewline
ndf & \multicolumn{3}{c|}{$3\times1\times2$} &
\multicolumn{3}{c|}{$3\times1\times2$} & $1\times3\times1$ & $1\times3\times1$ &
$1\times3\times1$ & $1\times3\times1$\tabularnewline
\hline 
\multicolumn{1}{c|}{Total (on-shell)} & \multicolumn{6}{c|}{12} &
\multicolumn{4}{c}{12}\tabularnewline
\hline
\end{tabular}
\end{table}

Naturally, the infinitisimal change in a scalar under a spin transformation will
require its right-hand side to be a scalar such that the spinor indices are
suppressed.
Also, it can be seen immediately that MUSY transformation
\ref{eq:MUSY-transformation}
should have a new parameter discussed previously to satisfy the family index in
both sides of the equation.
Therefore, the change in scalar takes the following form:

\begin{eqnarray}
\label{eq:scalar-transformation}
\begin{array}{rcl}
\delta\phi_{i} & = & \zeta_{i}^{a}\epsilon^{\alpha}\psi_{\alpha a}\\
\delta\hat{\phi}_{i} & = &
\zeta_{ia}\bar{\epsilon}_{\dot{\alpha}}\bar{\psi}^{\dot{\alpha}a}
\end{array}
\end{eqnarray}
where $\epsilon_{\alpha}$ is the so-called SUSY parameter, and $\zeta_{ia}$
are complex constants in a global MUSY transformation, i.e.
$\partial_{\mu}\zeta_{ia}=0$ \cite{Martin}.
One can note that 
\begin{eqnarray}
\begin{array}{rcl}
 \delta\phi^{*} & = & \delta\hat{\phi}\\
 \delta\hat{\phi}^{*} & = & \delta\phi
\end{array}
\end{eqnarray}
 although $\phi$ and $\hat{\phi}^{*}$ themselves are not equal (as
it should not), due to the condition of $n_{F}=n_{B}$. Now, we may deduce
the transformation of the fermionic preon for the free case by writing
the MUSY Lagrangian density, which will be discussed in the next section, as
follows:

\begin{eqnarray}
\mathcal{L}_{free} & = & -\partial^{\mu}\phi_{i}^{*}\partial_{\mu}\phi_{i}
-\partial^{\mu}\hat{\phi}_{i}^{*}\partial_{\mu}\hat{\phi}_{i}
+i\bar{\psi}^{a}\bar{\sigma}^{\mu}\partial_{\mu}\psi_{a}
\end{eqnarray}
where summation convention is used over repeated indices $i$, $a$ and
$\mu$. Therefore, the change in fermion field, which makes the action invariant
under
MUSY transformations up to total derivatives, is written as follows:
\begin{eqnarray}
  \delta\psi_{\alpha,a} & = &
-i\zeta_{ia}\sigma_{\alpha\dot{\beta}}^{\mu}\bar{\epsilon}^{\dot{\beta}}
\partial_{\mu}\left(\phi_{i}+\hat{\phi}_{i}^{*}\right) \label{eq:psi-degis}
\end{eqnarray}
 where $\sigma_{\mu}$ are the Pauli matrices.

In summary, the MUSY transformations take the following form, immediately after
taking the Hermitian conjugate of Eq. (\ref{eq:psi-degis}):
\begin{eqnarray}
\begin{array}{rcl}
\delta\phi_{i} & = & \bar{\zeta}_{i}^{a}\epsilon\psi_{a}\\
\delta\hat{\phi}_{i} & = & \zeta_{ia}\bar{\epsilon}\bar{\psi}^{a}\\
\delta\psi_{a} & = &
-i\zeta_{ia}\sigma^{\mu}\bar{\epsilon}\partial_{\mu}\left(\phi_{i}+\hat{\phi}_{i
}^{*}\right) \\
\delta\bar{\psi}^{a} & = &
i\bar{\zeta}_{i}^{a}\epsilon\sigma^{\mu}\partial_{\mu}\left(\phi_{i}^{*}+\hat{
\phi}_{i}\right)
\end{array}
\end{eqnarray}
 where the spinor indices are suppressed. This transformation indeed looks like
SUSY transformation of Wess-Zumino model, only with an additional colour
transformation.

\subsection{Effective MUSY Lagrangian\label{sec:effective-lagrangian}}

The indices of fermionic preons $\psi_{a}$, and scalars $\phi_{i}$
and $\hat{\phi_{i}}$ do not indicate colour and family \emph{a priori}.
Instead, when the gauge bosons emerges with an approximate gauge symmetry
from the current-current interactions as discussed in \cite{suzuki},
preons will have colour or flavour \emph{post-priori}. 
The simplest effective Lagrangian containing free fields and current-current
interactions
at preonic level can be written as follows (again suppressing the family and
colour indices):

\begin{eqnarray}
\mathcal{L}_{preon} & = & -\partial^{\mu}\Phi^{\dagger}\partial_{\mu}\Phi
+i\psi^{\dagger}\bar{\sigma}^{\mu}\partial_{\mu}\psi
+i\bar{\psi}^{\dagger}\sigma^{\mu}\partial_{\mu}\bar{\psi}+f_{s}^{2}j_{m}^{\mu}
j_{m\mu}
+f_{W}^{2}J_{k}^{\mu}J_{k\mu}+f^{\prime2}J^{\mu}J_{\mu} + f_0^2 j^{\mu} j_{\mu}
\label{eq:preonic-lagrangian}
\end{eqnarray}

where $\Phi$ is a doublet consist of the scalar preons, and $f_{s}$, $f_0$,
$f_{W}$ and $f^{\prime}$ are constants of dimension mass$^{-1}$, $j_{m}^{\mu}$,
$j^\mu$,
$J_{k}^{\mu}$ and $J^{\mu}$ are the currents of the preons which
can be explicitly written as:

\begin{eqnarray}
\begin{array}{rcl}
j_{m}^{\mu} & = &
\psi^{\dagger}\lambda_{m}\bar{\sigma}^{\mu}\psi+\bar{\psi}^{\dagger}\lambda_{m}
\sigma^{\mu}\bar{\psi} ,\\
j^{\mu} & = &
\psi^{\dagger}\lambda_{m}\bar{\sigma}^{\mu}\psi+\bar{\psi}^{\dagger}
\sigma^{\mu}\bar{\psi} ,\\
J_{k}^{\mu} & = &
\Phi^{\dagger}\tau_{k}\partial^{\mu}\Phi+(\partial^{\mu}\Phi^{\dagger})\tau_{k}
\Phi , \\
J^{\mu} & = &
\Phi^{\dagger}\partial^{\mu}\Phi+(\partial^{\mu}\Phi^{\dagger})\Phi .
\end{array}
\end{eqnarray}
Here $m=1,\cdots,8$ and $k=1,2,3$ indicates the indices of the generators
$\lambda_{m}$ and $\tau_{k}$ of groups $SU(3)$ and $SU(2)$, respectively.

One could easily see that Eq. (\ref{eq:preonic-lagrangian}) is a
non-renormalizable Lagrangian including a current-current interaction
and it is not invariant under MU transformations anymore. This non-invariance
occurs due to the current-current interaction, and it could be cured
by adding an auxiliary Lagrangian to $\mathcal{L}_{preon}$. The auxiliary
Lagrangian 
could be written as:

\begin{eqnarray}
\mathcal{L}_{aux} & = &
\left(m^{\prime}B^{\mu}-f^{\prime}J^{\mu}\right)\left(m^{\prime} B^{\mu}-f^{\prime}J^{\mu}
\right)+\left(m_{W}W_{k}^{\mu}-f_{W}J_{k}^{\mu}\right)\left(m_{W}W_{k\mu}-f_{W}
J_{k\mu}\right)\\
 & + &
\left(m_{s}G_{m}^{\mu}-f_{s}j_{m}^{\mu}\right)\left(m_{s}G_{m}^{\mu}-f_{s}j_{m}^
{\mu}\right) + \left(m_{0} G_0^{\mu}-f_{0} j^{\mu}\right) \left(m_{0}G_{0}^{\mu}
- f_{0}j^
{\mu}\right) \nonumber 
\end{eqnarray}
where the new fields, $G_{m}^{\mu}$, $G_0^\mu$, $W_{k}^{\mu}$ and $B^{\mu}$ are
the, so
called, auxiliary fields.
Then the two Lagrangians $\mathcal{L_{\mbox{preon}}}$ and
$\mathcal{L}=\mathcal{L_{\mbox{preon}}}+\mathcal{L_{\mbox{aux}}}$ are equivalent
since their
partition functions are equal. 
A general approach to preonic models where currents act as approximate gauge
fields, were investigated elsewhere \cite{suzuki}.

When off-shell, the new fields give an extra degree of freedom and they
should satisfy the MU invariance as these also transform under MU
transformation.
This requirement implies an additional term in the MU transformations of the
preons as follows:

\begin{eqnarray}
\begin{array}{rcl}
\delta\phi_{i} & = & \bar{\zeta}_{i}^{a}\epsilon\psi_{a}\\
\delta\hat{\phi}_{i} & = & \zeta_{ia}\bar{\epsilon}\bar{\psi}^{a}\\
\delta\psi_{a} & = &
-i\zeta_{ia}\sigma^{\mu}\bar{\epsilon} D_{\mu}\left(\phi_{i}+\hat{\phi}_{i
}^{*}\right)\label{eq:psi-degis-1}\\
\delta\bar{\psi}^{a} & = &
i\bar{\zeta}_{i}^{a}\epsilon\sigma^{\mu} \left[
D_{\mu}\left(\phi_{i}+\hat{\phi}_{i
}^{*}\right) \right]^{\dag}
\end{array}
\end{eqnarray}
where $D_\mu \Phi = \partial_\mu \Phi - i g^\prime Y^\prime B_\mu \Phi - i g_W
T_k W_\mu^k \Phi $
is the covariant derivatives are presented via the occurance of the approximate
gauge symmetry as motivated.

Since auxiliary fields do not have kinetic terms, the variation of the total
Lagrangian will give equations of motion of these fields
as follows:

\begin{eqnarray}
\begin{array}{rcl}
G_{m}^{\mu} & = &
\left(f_{s}/m_{s}\right)\left(\psi^{\dagger}\lambda_{m}\bar{\sigma}^{\mu}
\psi+\bar{\psi}^{\dagger}\lambda_{m}\sigma^{\mu}\bar{\psi}\right)\\
G_{0}^{\mu} & = &
\left(f_{s}/m_{s}\right)\left(\psi^{\dagger}\bar{\sigma}^{\mu}
\psi+\bar{\psi}^{\dagger}\sigma^{\mu}\bar{\psi}\right)\\
W_{k}^{\mu} & = &
\left(f_{W}/m_{W}\right)\left(\Phi^{\dagger}\tau_{k}\partial^{\mu}
\Phi+(\partial^{\mu}\Phi^{\dagger})\tau_{k}\Phi\right)\\
B^{\mu} & = &
\left(f^{\prime}/m^{\prime}\right)\left(\Phi^{\dagger}\partial^{\mu}
\Phi+(\partial^{\mu}\Phi^{\dagger})\Phi\right)
\end{array}
\end{eqnarray}
The coupling constants and masses of bosons obtained from loop diagrams
will give rise to an approximate $SU(3)\times SU(2)\times U(1)$ gauge
symmetry, with an extra singlet, where all the gauge non-invariant parts are
embedded into
the mass term of the vector fields: \begin{eqnarray}
\mathcal{L} & = &
-\partial^{\mu}\Phi^{\dagger}\partial_{\mu}\Phi+i\psi^{\dagger}\bar{\sigma}^{\mu
}\partial_{\mu}\psi+i\bar{\psi}^{\dagger}\sigma^{\mu}\partial_{\mu}\bar{\psi}+m^
{\prime2}B^{\mu}B_{\mu}+m_0^2 G_0^{\mu}G_{0\mu}\\
 & + &
m_{W}^{2}W_{k}^{\mu}W_{k\mu}+m_{s}^{2}G_{m}^{\mu}G_{m\mu}+g_{s}j_{m}^{\mu}G_{
m\mu}+g_{W}j_{k}^{\mu}W_{k\mu}+g_0 j_{0}^{\mu}G_{0\mu}\nonumber
\end{eqnarray}
where $g_s$, $g_0$, $g^{\prime}$ and $g_W$ are dimensionless coupling constants
satisfying
\begin{eqnarray}
\label{eq:coupling-constants}
\begin{array}{rcl}
 g_{s} &=& f_{s}m_{s} \\ 
 g_{0} &=& f_{0}m_{0} \\ 
 g_{W} &=& f_{W}m_{W} \\
 g^{\prime} &=& f^{\prime}m^{\prime}
\end{array}
\end{eqnarray}

The mass of the vector bosons $G^{\mu}$ and $G_0^\mu$
 converges to zero as the corresponding constant $f$ diverges
and the masses of weak bosons arise from the spontaneous symmetry
breaking as will be discussed in the next section. The procedure for obtaining
mass to composite vector bosons from loop diagrams
has already been investigated in detail \cite{suzuki}. The corresponding
representations
for the composite vector bosons are presented in Table \ref{Tab:reps}.

\begin{table}[ht]
\caption{Composite SM vector bosons
\label{Tab:reps}}
\begin{tabular}{l|ccc||cc}
Vector boson & Constituent & Representation \tabularnewline
\hline
\hline 
$G_m^\mu$ & $\bar\Psi \gamma \Psi$  & \textbf{8} \tabularnewline
$G_0^\mu$ & $\bar\Psi \gamma \Psi$  & \textbf{1} \tabularnewline
$W_k^\mu$ & $\Phi^\dagger \partial \Phi$  & \textbf{3} \tabularnewline
$B^\mu$ & $\Phi^\dagger \partial \Phi$  & \textbf{1}\tabularnewline
\hline
\end{tabular}
\end{table}

\section{Standard Model Particles in MUSY Model\label{sec:SMparticles}}

In this section quarks, leptons and gauge bosons will be defined in
the flavour basis. The corresponding states in the mass basis, resulting
in CKM mixings, are discussed in the next section. The SM fermions
are proposed as doublets made from fermionic and bosonic preons. Similarly,
SM gauge bosons are assumed to be matrices acting on those doublets.

\subsection{Standard Model fermions in MUSY model}

The SM fermions are in the form:

\begin{eqnarray*}
Q_{i} & = & \left(\begin{array}{c}
u\\
d\end{array}\right),\ \left(\begin{array}{c}
c\\
s\end{array}\right),\ \left(\begin{array}{c}
t\\
b\end{array}\right),\\
\bar{u}_{i} & = & \bar{u},\ \bar{c},\ \bar{t},\\
\bar{d}_{i} & = & \bar{d},\ \bar{s},\ \bar{b},\\
L_{i} & = & \left(\begin{array}{c}
\nu_{e}\\
e\end{array}\right),\ \left(\begin{array}{c}
\nu_{\mu}\\
\mu\end{array}\right),\ \left(\begin{array}{c}
\nu_{\tau}\\
\tau\end{array}\right),\\
\bar{e}_{i} & = & \bar{e},\ \bar{\mu},\ \bar{\tau},\\
\bar{\nu}_{i} & = & \bar{\nu}_{e},\ \bar{\nu}_{\mu},\
\bar{\nu}_{\tau},\end{eqnarray*}
where $Q_i$ and $L_i$ are weak isospin doublets and the remaining ones are
singlets.
In order to construct SM fermions in terms of MUSY, one can write the SM
particles
as composite states of preons. A quark is then considered as a composite state
made
of a fermionic preon and a scalar preon. Thus, it has colour and
fractional electric charge as it should. The preonic scalar is to be postulated
as $\phi_{i}$
for up-type and $\hat{\phi}_{i}$ for down-type quarks where $i=1,2,3$
denotes the quark family index. Similarly, up and down type antiquarks
are in the form $\bar{\psi}\phi_{i}^{*}$ and $\bar{\psi}\hat{\phi}_{i}^{*}$,
respectively. The full preon content of all left and
right handed quark states can be found in Table \ref{Tab:quarks-musy}.
Note that we have written the MU numbers of composite particles by
summing MU numbers of the preons. 

\begin{table}[ht]
\caption{SM quarks according to the MUSY model in flavour basis
\label{Tab:quarks-musy}}
\begin{tabular}{l|ccc||cc}
Quark (left) & Q  & C  & MU$_{L}$  & Quark (right) & MU$_{R}$ \tabularnewline
\hline
\hline 
$u=\psi\phi_{1}$ & $2/3$  & \textbf{3}  & 1 & $\bar{u}=\bar{\psi}\phi_{1}^{*}$ &
0\tabularnewline
$d=\psi\hat{\phi}_{1}$ & $-1/3$  & \textbf{3}  & 0 &
$\bar{d}=\bar{\psi}\hat{\phi}_{1}^{*}$ & -1\tabularnewline
$c=\psi\phi_{2}$ & $2/3$  & \textbf{3}  & 1 & $\bar{c}=\bar{\psi}\phi_{2}^{*}$ &
0\tabularnewline
$s=\psi\hat{\phi}_{2}$ & $-1/3$  & \textbf{3}  & 0 &
$\bar{s}=\bar{\psi}\hat{\phi}_{2}^{*}$ & -1\tabularnewline
$t=\psi\phi_{3}$ & $2/3$  & \textbf{3}  & 1 & $\bar{t}=\bar{\psi}\phi_{3}^{*}$ &
0\tabularnewline
$b=\psi\hat{\phi}_{3}$ & $-1/3$  & \textbf{3}  & 0 &
$\bar{b}=\bar{\psi}\hat{\phi}_{3}^{*}$ & -1\tabularnewline
\hline
\end{tabular}
\end{table}

Leptons are thought to be composed of three fermionic preons and one
bosonic preon as listed in Table \ref{tab:leptons_musy}. To keep compatibility
with the SM spin 1/2 particles,
one fermionic preon will be opposite handed with respect to the other
two. In terms of the colour structure, the state counting gives
$\mathbf{3}\otimes\mathbf{3}\otimes\mathbf{3}=\mathbf{1}\oplus\mathbf{8}
\oplus\mathbf{8}\oplus\mathbf{10}$
states. The colour singlet state is taken to yield the SM leptons.
The remaining states will be discussed later in Section \ref{sec:Oghuz}.

The masses of the SM fermions could arise from the quantum mass corrections
from the four-point interaction of the fermionic preons, similar to the mass
term in $\phi^4$-model \cite{ryder}.
Fermionic component of the bound state resulting in SM particles could also help
acquiring mass through (approximate) guage interactions even if
the fermionic preon does not interact directly with the scalar one
\cite{Martin}.
In general, the correction to the squared mass of the scalar preon has two
leading terms one from self-interaction and another one from an indirect
interaction with the fermionic preon. However, since the self-interaction term does not
exist due to the absence of the Yukawa coupling as MU symmetry requires, the correction to the squared mass of the
scalar preon would be as follows:

\begin{eqnarray}
\Delta M_{\phi}^{2} & = &
-\frac{g_s^{2}+g_0^{2}}{16\pi^{2}}\left[\Lambda^{2}-24 M_{\psi}^{
2}\ln\left(\Lambda/M_{\psi}\right)+\cdots\right]\end{eqnarray}
where $\Lambda$ is the cut-off scale and it gives contribution
even if the bare mass $M_{\psi}$ is set to zero. The mass terms will
be finite since $f_s$ and $f_0$ in Eq. (\ref{eq:coupling-constants}) are expected
to be in the same order
of magnitude. At higher energies, the contribution from the fermionic
preons is suppressed since there aren't any direct interactions between
fermionic and scalar preons. However, the masses are still finite since
$f^{2}$ and $1/\Lambda^{2}$ terms are expected to be at the same order of
magnitude. This can be
understood by considering the preonic four-fermion interaction constant $f^2$ as
an analogy
to the Fermi constant $G_F = g^2_W / M_W^2$ which is in the same order of
magnitude with $1/\Lambda_{\mbox{weak}}^2$ 
where $\Lambda_{\mbox{weak}} \approx M_W$ is the cut-off scale of the Fermi
theory.

According to the discussion above, the masses of the composite states could be obtained
without having an hierarchy problem and furthermore the composite states would acquire their masses
with respect to the chiral symmetry breaking of the effective action similar to the baryons in
chiral perturbation theory \cite{Ecker:1994gg}.

\begin{table}[ht]
\caption{SM leptons according to the MUSY model in Flavour Basis
\label{tab:leptons_musy}}
\begin{tabular}{l|ccc||cc}
 Left handed states & Q  & C  & MU$_{L}$  &  Right handed states & MU$_{R}$
\tabularnewline
\hline
\hline 
$\nu_{e}=\psi_{R}\bar{\psi}_{G}\psi_{B}\phi_{1}$  & 0  & \textbf{1}  & 1 &
$\bar{\nu}_{e}=\bar{\psi}_{R}\psi_{G}\bar{\psi}_{B}\phi_{1}^{*}$  &
0\tabularnewline
$e=\psi_{R}\bar{\psi}_{G}\psi_{B}\hat{\phi}_{1}$  & -1  & \textbf{1}  & 0 &
$\bar{e}=\bar{\psi}_{R}\psi_{G}\bar{\psi}_{B}\hat{\phi}_{1}^{*}$  &
-1\tabularnewline
$\nu_{\mu}=\psi_{R}\bar{\psi}_{G}\psi_{B}\phi_{2}$  & 0  & \textbf{1}  & 1 &
$\bar{\nu}_{\mu}=\bar{\psi}_{R}\psi_{G}\bar{\psi}_{B}\phi_{2}^{*}$  &
0\tabularnewline
$\mu=\psi_{R}\bar{\psi}_{G}\psi_{B}\hat{\phi}_{2}$ & -1  & \textbf{1}  & 0 &
$\bar{\mu}=\bar{\psi}_{R}\psi_{G}\bar{\psi}_{B}\hat{\phi}_{2}^{*}$  &
-1\tabularnewline
$\nu_{\tau}=\psi_{R}\bar{\psi}_{G}\psi_{B}\phi_{3}$  & 0  & \textbf{1}  & 1 &
$\bar{\nu}_{\tau}=\bar{\psi}_{R}\psi_{G}\bar{\psi}_{B}\phi_{3}^{*}$  &
0\tabularnewline
$\tau=\psi_{R}\bar{\psi}_{G}\psi_{B}\hat{\phi}_{3}$  & -1  & \textbf{1}  & 0 &
$\bar{\tau}=\bar{\psi}_{R}\psi_{G}\bar{\psi}_{B}\hat{\phi}_{3}^{*}$  &
-1\tabularnewline
\hline
\end{tabular}
\end{table}

\subsection{Standard Model gauge bosons in MUSY model}

At the preonic scale, MUSY model does not have fundamental gauge interactions
at all. This means that a preon does not decay into another preon.
Since SM consists of the gauge group $SU(3)\times SU(2)\times U(1)$,
MUSY model should produce those gauge bosons, out of the preons in
the model. 

One way of forming the gauge bosons, similar to the construction of the scalar
bosons in Technicolour models, is to
form a fermionic bound state. However, one could easily see that fermionic
composite gauge bosons would have to be flavour singlets since the
flavour is originated only from bosonic preons in MUSY model. Another
way is current-current interactions becoming current-gauge interactions
of SM fermions and thus all terms that are not invariant under global
gauge transformations transferred into boson mass \cite{suzuki}.
Fermionic (left and right) and bosonic current definitions are repeated
here for the reader's convenience:

\begin{eqnarray}
\begin{array}{rcl}
G_{m}^{\mu} & = &
g_{s}\left(\psi^{\dagger}\lambda_{m}\bar{\sigma}^{\mu}\psi+\bar{\psi}^{\dagger}
\lambda_{m}\sigma^{\mu}\bar{\psi}\right)\\
G_{m}^{\mu} & = &
g_{0}\left(\psi^{\dagger} \bar{\sigma}^{\mu}\psi+\bar{\psi}^{\dagger}
\lambda_{m}\sigma^{\mu}\bar{\psi}\right)\\
W_{k}^{\mu} & = &
g_{W}\left(\Phi^{\dagger}\tau_{k}\partial^{\mu}\Phi+(\partial^{\mu}\Phi^{\dagger
})\tau_{k}\Phi\right)\\
B^{\mu} & = &
g^{\prime}\left(\Phi^{\dagger}\partial^{\mu}\Phi+(\partial^{\mu}\Phi^{\dagger}
)\Phi\right)
\end{array}
\end{eqnarray}
where the fields $\psi$, $\bar{\psi}$ and $\Phi$ are multiplets
of the groups generated by $\lambda_{m}$ and $\tau_{n}$, respectively.

\subsubsection{Gluons}

The fermionic preon current gives a colour-changing vector gauge boson
which is nothing but the gluon. Since the multiplets in the current
are colour triplets, the gauge group is nothing but the $SU(3)_{\text{colour}}$.
Therefore, gluons are defined proportional to the colour-changing
current which is unique in the model: \begin{equation}
G_{m}^{\mu}\sim\bar{\Psi}\gamma^{\mu}\Psi\end{equation}
where $\lambda_{m}$ are the Gell-Man matrices for $m=1,\ \cdots,\ 8$
and $\bar{\Psi}$ refers to a four-component fermion
taking $\psi$ and $\bar{\psi}$ as left and right handed colour triplet
fermions, respectively. Table \ref{tab:gluons_in_musy} contains
a summary of the preonic contents of the gluons according to the MUSY
model. The mass of the gluons emerges as approximately zero since
the four-point interaction of the fermionic preons is infinitely strong,
i.e., $m_{s}\rightarrow0$ as $f_{s}\rightarrow\infty$ since $g_{s}$
is known and finite from the experiments. It is crucial to have a sufficiently
large $f_s^2$ because the preonic four-fermion interaction is expected to have
a very large cut-off scale with respect to the scalar current-current
interactions.
This cut-off scale is the cut-off scale of the effective Lagrangian discussed in
Section \ref{sec:effective-lagrangian}. The same approach is applied to the
field
$G_0^\mu$ which is a colour singlet and interacts only with fermionic preons.

\begin{table}[ht]
\caption{SM Gluons according to the MUSY model. Here the indices of fermionic
preons refers R for red, B for blue and G for green. \label{tab:gluons_in_musy}}
\begin{tabular}{l|c}
Gluon  & Preonic Contents \tabularnewline
\hline
\hline 
$G_{1}^{\mu}$  &
$\frac{1}{\sqrt{2}}\left(\bar{\Psi}^{R}\gamma^{\mu}\Psi_{B}+\bar{\Psi}^{B}
\gamma^{\mu}\Psi_{R}\right)$ \tabularnewline
$G_{2}^{\mu}$  &
$\frac{-i}{\sqrt{2}}\left(\bar{\Psi}^{R}\gamma^{\mu}\Psi_{B}-\bar{\Psi}^{B}
\gamma^{\mu}\Psi_{R}\right)$ \tabularnewline
$G_{3}^{\mu}$  &
$\frac{1}{\sqrt{2}}\left(\bar{\Psi}^{R}\gamma^{\mu}\Psi_{R}-\bar{\Psi}^{B}
\gamma^{\mu}\Psi_{B}\right)$ \tabularnewline
$G_{4}^{\mu}$  &
$\frac{1}{\sqrt{2}}\left(\bar{\Psi}^{R}\gamma^{\mu}\Psi_{G}+\bar{\Psi}^{G}
\gamma^{\mu}\Psi_{R}\right)$ \tabularnewline
$G_{5}^{\mu}$  &
$\frac{-i}{\sqrt{2}}\left(\bar{\Psi}^{R}\gamma^{\mu}\Psi_{G}-\bar{\Psi}^{G}
\gamma^{\mu}\Psi_{R}\right)$\tabularnewline
$G_{6}^{\mu}$  &
$\frac{1}{\sqrt{2}}\left(\bar{\Psi}^{B}\gamma^{\mu}\Psi_{G}+\bar{\Psi}^{G}
\gamma^{\mu}\Psi_{B}\right)$ \tabularnewline
$G_{7}^{\mu}$  &
$\frac{i}{\sqrt{2}}\left(\bar{\Psi}^{B}\gamma^{\mu}\Psi_{G}+\bar{\Psi}^{G}
\gamma^{\mu}\Psi_{B}\right)$ \tabularnewline
$G_{8}^{\mu}$  &
$\frac{1}{\sqrt{6}}\left(\bar{\Psi}^{R}\gamma^{\mu}\Psi_{R}+\bar{\Psi}^{G}
\gamma^{\mu}\Psi_{G}-2\bar{\Psi}^{B}\gamma^{\mu}\Psi_{B}\right)$ \tabularnewline
\hline
$G_{0}^{\mu}$  &
$\left(\bar{\Psi}^{R}\gamma^{\mu}\Psi_{R}+\bar{\Psi}^{G}
\gamma^{\mu}\Psi_{G}+\bar{\Psi}^{B}\gamma^{\mu}\Psi_{B}\right)$ \tabularnewline
\hline
\end{tabular}
\end{table}

\subsubsection{Electroweak bosons and weak mixing angle}

The definition of the weak isotriplet and isosinglet fields emerge
from the MUSY model as bosonic currents with $SU(2)\times U(1)$ generators
since the only available preons are $\phi$ and $\hat{\phi}$,
corresponding to up and down isospin states. The explicit form of
the electroweak gauge bosons are shown in Table \ref{tab:EWbosons}.

\begin{table}[ht]
\caption{Electroweak Gauge Fields according to the MUSY model 
\label{tab:EWbosons}}
\begin{tabular}{c|c}
\noalign{\vskip\doublerulesep}
Fields  & Contents \tabularnewline[\doublerulesep]
\hline
\hline 
$W_{1}^{\mu}$  &
$\hat{\phi}^{\dagger}\partial^{\mu}\phi+\phi^{\dagger}\partial^{\mu}\hat{\phi}
$\tabularnewline[\doublerulesep]
$W_{2}^{\mu}$  &
$i\left(\hat{\phi}^{\dagger}\partial^{\mu}\phi-\phi^{\dagger}\partial^{\mu}\hat{
\phi}\right)$\tabularnewline[\doublerulesep]
$W_{3}^{\mu}$  &
$\phi^{\dagger}\partial^{\mu}\phi-\hat{\phi}^{\dagger}\partial^{\mu}\hat{\phi}
$\tabularnewline[\doublerulesep]
$B^{\mu}$  &
$\phi^{\dagger}\partial^{\mu}\phi+\hat{\phi}^{\dagger}\partial^{\mu}\hat{\phi}
$\tabularnewline[\doublerulesep]
\hline
\end{tabular}
\end{table}

The explicit preon content of the weak and electromagnetic gauge bosons are
given
in Table \ref{tab:WZA}. The masses of the weak bosons are obtained
via the mixing of the weak currents. With the well-known
mixing mechanism as introduced in \cite{FriztschM-1}, it is possible
to obtain a massless photon and masses of $W$ and $Z$ bosons compatible
with the experimental results. Following the electroweak model, the physical
states of charged gauge bosons ($W^{+}$ and $W^{-}$) can be defined as:
\begin{eqnarray}
W^{\pm\mu} & = & (W_{1}^{\mu}\mp iW_{2}^{\mu})/\sqrt{2}
\end{eqnarray}
 and the neutral bosons ($Z^{0}$ and photon) are defined through
the weak mixing angle: 
\begin{eqnarray}
\label{eq:mixSM}
\begin{array}{rcl}
Z^{\mu} & = & \cos\theta_{W}W_{3}^{\mu}-\sin\theta_{W}B^{\mu}\,, \\
A^{\mu} & = &
\sin\theta_{W}B^{\mu}+\cos\theta_{W}W_{3}^{\mu}\,.
\end{array}
\end{eqnarray}
Before the electroweak mixing, the masses of $W_1$, $W_2$ and $W_{3}$ are
equal as discussed in Ref. \cite{FriztschM-1}. However, when $W^{3}$ and $B$ are
mixed
to form the $Z$ boson and the physical photon as in Eq. (\ref{eq:mixSM}),
the contribution to the mass of $Z$ boson will be related
to the decay constant $F_{3}$ to the constituents in the vector boson,
which reads as follows:

\begin{eqnarray}
\langle0|\ J^{3\mu}\ |W^{3}\rangle & = & \langle0|\
\phi^{\dagger}\partial^{\mu}\phi-\hat{\phi}^{\dagger}\partial^{\mu}\hat{\phi}\
|W^{3}\rangle\\
 & = & iM_{w} F_{3} p^{\mu}
\end{eqnarray}
where $M_{W}$ is the mass of the $W^{3}$ boson (before the contribution),
$p^{\mu}$ is the momentum. Therefore, the mixing parameter (as in
\cite{FriztschM-1}) becomes $\lambda=\sqrt{g/e}\ \sin\theta_{W}=gM_{W}/F_{3}$.
Therefore the relation between the masses of the $Z$ and $W$ bosons can be
written as:
\begin{eqnarray}
M_{Z}^{2} & = & M_{W}^{2}/(1-\lambda^{2}) \, .
\end{eqnarray}
This mechanism is an anology of the mixing of the vector mesons of QCD
and obtaining their masses.

\begin{table}[ht]
\caption{Weak and Electromagnetic Gauge Fields according to the MUSY model where
$g_1=\cos \theta_W + \sin \theta_W$ and $g_2 = \cos \theta_W - \sin \theta_W$. 
\label{tab:WZA}}
\begin{tabular}{c|c}
\noalign{\vskip\doublerulesep}
Fields  & Contents \tabularnewline[\doublerulesep]
\hline
\hline 
$W_{\mu}^{+}$ &
$\hat{\phi}_{j}^{\dagger}\partial_{\mu}\phi_{i}$\tabularnewline
$W_{\mu}^{-}$ &
$\phi_{i}^{\dagger}\partial_{\mu}\hat{\phi}_{j}$\tabularnewline
$Z_{\mu}$ &
$\frac{1}{\sqrt{2}}\left(g_{1}\phi_{i}^{\dagger}\partial_{\mu}\phi_{i}-g_{2}\hat
{\phi}_{j}^{\dagger}\partial_{\mu}\hat{\phi}_{j}\right)$\tabularnewline
$A_{\mu}$ &
$\frac{1}{\sqrt{2}}\left(g_{2}\phi_{i}^{\dagger}\partial_{\mu}\phi_{i}+g_{1}\hat
{\phi}_{j}^{\dagger}\partial_{\mu}\hat{\phi}_{j}\right)$\tabularnewline
\hline
\end{tabular}
\end{table}

\section{SM Phenomenology within MUSY \label{sec:fenomenoloji}}

\subsection{Baryon and Lepton number conservation}

Let $\Delta_{\psi}$ indicate the difference between the number of
fermionic preons and fermionic anti-preons, and similarly $\Delta_{\phi}$
the difference between the number of scalars and anti-scalars. Since
a lepton includes three fermionic preons and one scalar preon, lepton
number is defined as:
\begin{eqnarray}
L & = & N_{\ell}-N_{\bar{\ell}}=\frac{\Delta\psi-\Delta\phi}{2}\,.
\end{eqnarray}
 As for quarks, there is one fermionic preon for each scalar. Thus,
one can define baryon number as follows:
\begin{eqnarray}
B & = & \frac{N_{q}-N_{\bar{q}}}{3}=\frac{3\Delta_{\phi}-\Delta_{\psi}}{6}\,.
\end{eqnarray}
 Since, there is no decay between fermionic preons and scalars in
MUSY model, these quantum numbers are always conserved. Hence, baryon
number violations such as proton decays is not allowed in MUSY model.

\subsection{Beta decay}

\begin{figure}
\includegraphics[scale=1.2]{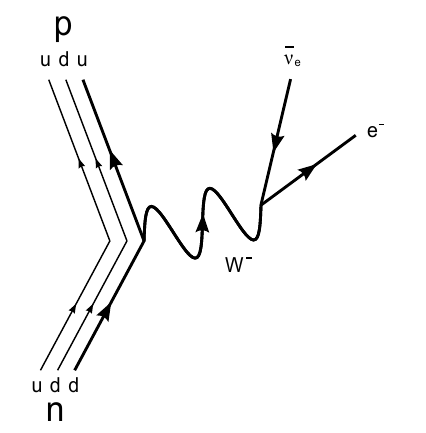}
$\quad$\includegraphics[scale=0.6]{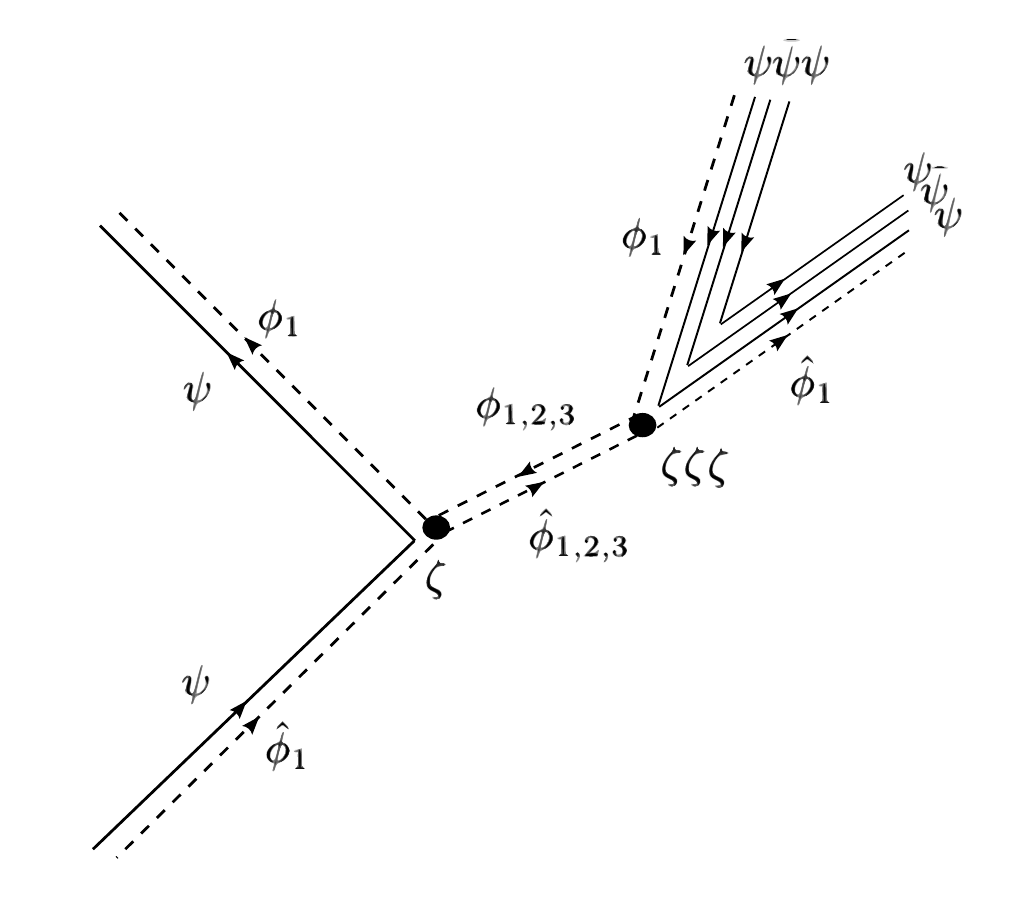}
\caption{Beta Decay: on the left according to the SM and on the right according
to MUSY model \label{fig:SMbeta}}
\end{figure}

Remembering the MUSY content of the quarks as $u=\psi\phi_{1}$ and
$d=\psi\hat{\phi}_{1}$ where the fermions are of the same colour,
the decay can be written as \begin{eqnarray}
\psi\hat{\phi}_{1} & \rightarrow & \psi\phi_{1}+W^{-}\,.\end{eqnarray}
It is already introduced that the charged weak boson is originating
from the current of scalar preons leading to an interaction
of current and vector boson just as in SM. The decay would consist
of four fermion and four scalar vertices at the preonic
level as shown in the Figure \ref{fig:SMbeta}.

\subsection{CKM and MNS mixing}

From some decay processes, it is known that quarks are observed in
mixed flavour while leptons remain in flavour basis. One could suggest
that this is because of the nature of quarks but not leptons. However,
MUSY model gives a different mixing mechanism which is originated
from the MUSY parameter $\zeta$ and the mixing is generated by the
weak bosons. This can be seen immediately by writing the low-energy
approximation
of the effective MUSY Lagrangian since the operators obtained would include
$\zeta_{ia}$ parameter as Noether current in Eq. (\ref{eq:noether_current})
arises. 
In order to have the mixing correct which the values would be obtained from
experiments,
the leptonic vertex would have $\zeta^3$ where the deviation from the unitarity
is
in the order of MNS mixing.

\section{Predictions of the MUSY model\label{sec:Predictions-of-MUSY}}

\subsection{Excited fermions}

As previously discussed, SM fermions are constructed as bound states
of preons: one scalar and one fermionic preon for quarks and one scalar
and three fermionic preons for leptons. Therefore, the MUSY model
contains a rich phenomenology due to the orbital or spin excitations
of the SM fermions.
For example, for each quark and lepton one expects to have a multitude
of orbital excitations, denoted as $q^{(n)}$ and $\ell^{(n)}$. In
the lepton sector, an additional possibility are the spin excitations
of the from $\ell_{3/2}\equiv\psi\psi\psi\phi$, giving spin 3/2 particles
and naturally their orbital excitations as well. So,
each SM lepton has two colour-octet and one colour-decouplet partners.

\subsection{New matter particles: The Oghuz \label{sec:Oghuz}}

As discussed in the construction of the SM leptons, three fermionic
preons yield a singlet, two octet and one decouplet
in terms of group representations of colours. It has been stated that
the singlet state corresponds to SM leptons, we therefore propose the remaining
objects as dark matter candidates with an appropriate helicity
state to give a spin 1/2 particle because there occurs chargeless
and colour-singlet fermions (like a spin-3/2 neutrino). 

\begin{table}[ht]
\caption{BSM particle content of the MUSY model \label{tab:oghuz}}
\begin{tabular}{l|ccc}
 Left handed states & Q  & C  & S\tabularnewline
\hline
\hline 
$\bar{\psi}\psi\phi$  & 1/3 & \textbf{$\mathbf{\bar{3}\oplus}\mathbf{6}$} &
0\tabularnewline
$\bar{\psi}\psi\hat{\phi}$  & -2/3 & \textbf{$\mathbf{\bar{3}\oplus}\mathbf{6}$}
& 0\tabularnewline
$\psi\psi\phi$  & 1/3 & \textbf{$\mathbf{\bar{3}\oplus}\mathbf{6}$} &
1\tabularnewline
$\psi\psi\hat{\phi}$  & -2/3 & \textbf{$\mathbf{\bar{3}\oplus}\mathbf{6}$} &
1\tabularnewline
$\psi\psi\psi\phi$  & 0  &
\textbf{$\mathbf{1}\oplus\mathbf{8}\oplus\mathbf{8}\oplus\mathbf{10}$} &
3/2\tabularnewline
$\psi\psi\psi\hat{\phi}$  & -1  &
\textbf{$\mathbf{1}\oplus\mathbf{8}\oplus\mathbf{8}\oplus\mathbf{10}$} &
3/2\tabularnewline
\hline
\end{tabular}
\end{table}

Possible bound states other than SM particles are in the form given
in Table \ref{tab:oghuz} with their charge, colour and spin properties.
These bound states are split into two subgroups which do not interact
with each other: a group of bicoloured bosonic (vector or scalar)
particles and another group of fermionic particles with 3/2 spin.
A suitable name these new particles could be ``Oghuz''%
\footnote{The name Oghuz is derived from the Turkish word ``ok'', which
means arrow or tribe. The reason for calling these particles as Oghuz
is simple: in Turkish history, the Oghuz tribes were split into two
fractions ``Üçoklar'' (Three Arrows) and ``Bozoklar'' (Grey
Arrows). These two fractions would not interact one with another,
just like the two sub-groups in the MUSY model.}, and the two sub-groups
could be called as ``Bozoklar'' and ``Üçoklar''
respectively. The Bozoklar interact within the group via four point
interactions consist of either fermionic preons and bosonic anti-preons
or fermionic anti-preons and bosonic preons. Therefore these ``particles''
can not interact with any of the SM bosons which consist of either
all preons or all anti-preons and can be though as the Dark Matter
(DM) candidates of the MUSY model. Additionally, these DM candidates
form a group of particles with two fermionic preons
and another group with either one or three fermionic preons.

\section{Discussion and conclusion}

A new symmetry, MU symmetry, including spin, colour and charge was
introduced. It was shown that the MU symmetry is a generalization
of the well known supersymmetry concept and in fact, reduces to SUSY
in a specific scenario. Using MU symmetry, and starting from a single
chiral colour triplet preon, the electroweak and QCD was built. The MUSY model
proposes SM fermions and gauge bosons as the bound states of scalar and
fermionic
preons. It also predicts the number of families as
three, relating it to the number of colours in strong interactions.
Other naturally occurring features of the model are the inclusion
of the lepton and baryon number conservation, the existence of the
quark mixings and new fermions which do not interact with the SM fermions.
These new fermions, named the Oghuz, could be the Dark Matter candidates.
The higher excitations of the SM bound states, are also expected to
be present in Nature and are to be sought at the present and future
colliders.

With an effective Lagrangian the MUSY model proposes the gauge bosons
of the SM as current-current interactions. It is also contains vanishingly
small gluon and photon masses due to infinitely strong
four-point interaction of the fermionic preons. The electroweak sector
remains very similar to the SM, including the weak mixing angle, arising
from four-point scalar preon interactions. Although the effective
Lagrangian approach does provide a working model, it would have been
more elegant to embed all preonic states in a large enough gauge group
similar to GUT models, i.e., a gauge interaction between preons an extension of
MUSY
model. The applicability of this idea is yet to be investigated.

\subsection*{Acknowledgements}
The authors would like to thank D. A. Demir, A. Havare and V. E. Ozcan for
useful
discussions.

\end{document}